\documentclass[sigconf]{acmart}
\usepackage{graphicx} 
\usepackage[]{algorithm2e} 
\usepackage{algorithmic} 
\AtBeginDocument{%
  }

\copyrightyear{2025} 
\acmYear{2025} 
\setcopyright{cc}
\setcctype{by}
\acmConference[E-ENERGY '25]{The 16th ACM International Conference on Future and Sustainable Energy Systems}{June 17--20, 2025}{Rotterdam, Netherlands} 
\acmBooktitle{The 16th ACM International Conference on Future and Sustainable Energy Systems (E-ENERGY '25), June 17--20, 2025, Rotterdam, Netherlands}
\acmDOI{10.1145/3679240.3734589}
\acmISBN{979-8-4007-1125-1/2025/06}

\begin{document}

\title{GenTL: A General Transfer Learning Model for Building Thermal Dynamics}

\author{Fabian Raisch}
\authornote{Both authors contributed equally to the research project.}
\authornote{Email: fabian.raisch@th-rosenheim.de; main affiliation at Technical University of Applied Sciences Rosenheim; doctoral candidate at Technical University of Munich (cooperative doctorate with Technical University of Applied Sciences Rosenheim).}
\orcid{1234-5678-9012}
\affiliation{%
  \institution{Technical University of Applied Sciences Rosenheim}
  \city{Rosenheim}
  \country{Germany}}
\affiliation{%
\institution{Technical University of Munich}
  \city{Munich}
  \country{Germany}}

\author{Thomas Krug}
\authornotemark[1]
\affiliation{%
  \institution{Technical University of Applied Sciences Rosenheim}
  \city{Rosenheim}
  \country{Germany}}
\additionalaffiliation{%
    \institution{Karlsruhe Institute of Technology}
  \city{Karlsruhe}
  \country{Germany}}

\author{Christoph Goebel}
\affiliation{%
  \institution{Technical University of Munich}
  \city{Munich}
  \country{Germany}}

\author{Benjamin Tischler}
\affiliation{%
 \institution{Technical University of Applied Sciences Rosenheim}
  \city{Rosenheim}
  \country{Germany}}

\renewcommand{\shortauthors}{Raisch et al.}

\begin{abstract}

Transfer learning (TL) is an emerging field in modeling building thermal dynamics. This method reduces the data required for a data-driven model of a target building by leveraging knowledge from a source building. Consequently, it enables the creation of data-efficient models that can be used for advanced control and fault detection \& diagnosis.
A major limitation of the TL approach is its inconsistent performance across different sources. 
Although accurate source-building selection for a target is crucial, it remains a persistent challenge.

We present GenTL, a general transfer learning model for single-family houses in Central Europe.
GenTL can be efficiently fine-tuned to a large variety of target buildings. It is pretrained on a Long Short-Term Memory (LSTM) network with data from 450 different buildings. The general transfer learning model eliminates the need for source-building selection by serving as a universal source for fine-tuning. 
Comparative analysis with conventional single-source to single-target TL demonstrates the efficacy and reliability of the general pretraining approach. Testing GenTL on 144 target buildings for fine-tuning reveals an average prediction error (RMSE) reduction of 42.1\% compared to fine-tuning single-source models.

\end{abstract}




\maketitle

\section{Introduction}

Data-driven models (DDM) of thermal dynamic systems in buildings become increasingly important since they enable advanced control for more energy-efficient operation as well as fault detection \& diagnosis (FDD).
The creation of such models requires a large amount of measurement data. For this reason, transfer learning (TL) is currently receiving a lot of attention. 
TL leverages knowledge from a pretrained DDM of a source building. The pretrained model is fine-tuned for a specific target building using data from the target. 
Typically, a relatively small amount of target data is required to achieve good prediction quality. Consequently, TL reduces the volume of data required from the target building. Furthermore, fine-tuned models often exhibit lower prediction errors compared to models trained from scratch using the same amount of limited data.
In TL for DDM, a distinction can be made between single-source (pretraining on one building) and multi-source (pretraining on several buildings). For demonstrating TL, single or multiple targets can be utilized.
Single-source to single-target TL is the most presented approach in TL studies applied to building thermal dynamics \cite{chen2020transfer, pinto2022transfer, jiang2019deep}. However, these studies have demonstrated TL exclusively for specific use cases. This limitation poses challenges in proving the general applicability of the TL principle.
As a result, recent studies have evaluated TL on multiple targets \cite{pinto2022sharing, li2024building}. 
Nevertheless, all studies in the building dynamics domain utilized only single sources for TL, limiting the scope of the observed scenarios. Therefore, \cite{pinto2022transfer, li2024building, jiang2019deep} request the use of multiple source buildings in the process of transferring knowledge. This raises the question of how a multi-source approach can effectively leverage knowledge from multiple sources for TL. 
In addition, it remains unclear whether a multi-source approach will result in lower prediction errors after fine-tuning compared to conventional single-source to single-target methods. 

Another challenge in the area of TL is the selection of an appropriate source model for a specific target, as assessing the similarity between buildings is complex. In \cite{li2024building}, the authors suggest random source selection within an archetype of buildings, also for the reason that an excessive search is computationally expensive. Consequently, another question arises: can a multi-source approach outperform random source selection in this context? 

To address these challenges, we introduce a general pretrained data-driven model for indoor temperature prediction. The general pretrained model is designed for single-family houses in Central Europe. We use synthetic data from 450 individual buildings generated with a Modelica simulation \cite{mattsson1997modelica}, exported as a Functional Mock-up Unit (FMU) \cite{blochwitz2012functional}, and executed in Python. This data is used for training a three-layer Long Short-Term Memory (LSTM) neural network \cite{schmidhuber1997long}. We introduce the first multi-source approach for TL in building thermal dynamics. The pretrained model is then fine-tuned on various target buildings and compared with conventional single-source to single-target models to determine whether multi-source or single-source TL yields lower prediction errors. Additionally, we investigate the performance of random source selection versus the general source model for fine-tuning. These findings help us recommend how pretrained models should be designed to achieve low prediction errors for broad applicability.

\begin{figure*}
    \centering
    \includegraphics[width=0.8 \linewidth]{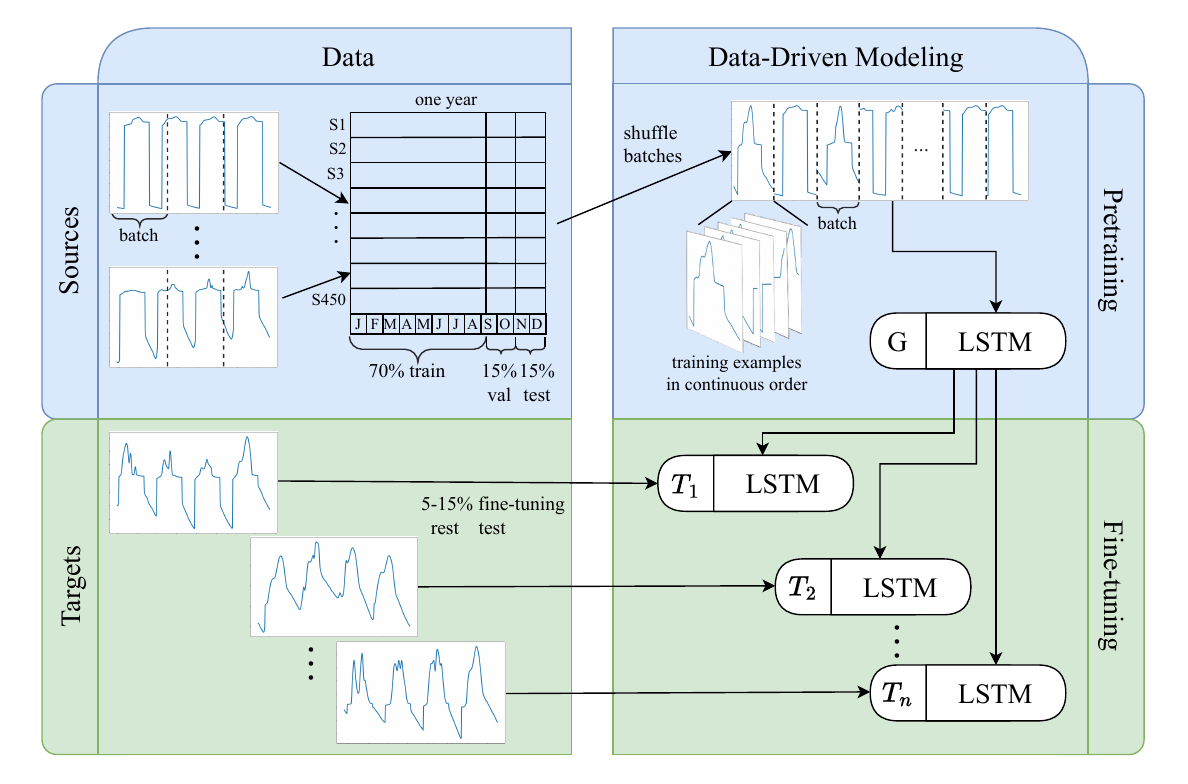}
    \caption{Overview of method for pretraining and fine-tuning the general source model G}
    \label{fig:MethodicalConcept}
\end{figure*}

\section{Related Work}
\label{ch:review}

DDM in the area of building thermal dynamics has been used various times in the literature. Next to classical DDM approaches as in \cite{chen2016fractional}, machine learning (ML)-based methods such as LSTMs are utilized to capture the thermal dynamics of buildings \cite{Elmaz2021CNNLSTM, Martinez2020LSTMMLP, Mtibaa2020LSTM}. 
A key application of DDMs in buildings is energy-efficient control, where they primarily serve as models for Model Predictive Control (MPC) \cite{Drgona.2020, stoffel2023evaluation}. Additionally, Reinforcement Learning (RL) agents can be utilized for control, with the agent pretrained on a DDM before deployment \cite{YU2022109458, CORACI2023117303}. Another important application of DDMs is Fault Detection \& Diagnosis (FDD), where they act as baselines to compare operational data and identify faults \cite{chen2023review}.

Due to limited data availability, TL is a promising method for DDM of buildings \cite{peirelinck2022transfer, pinto2022transfer}. In \cite{jiang2019deep}, for example, the authors used LSTMs with sequence-to-sequence modeling to simulate indoor thermal dynamics based on two public datasets. Similarly, the authors of \cite{grubinger2017generalized} presented an online approach that was further extended to an MPC controller. In \cite{chen2020transfer}, Chen et al. demonstrated the use of TL for indoor temperature and humidity prediction with natural ventilation for a source room in Beijing. For fine-tuning, they used 15 consecutive days of data in a target room in Shanghai with different building sizes, construction materials, and glazing areas. 
These studies generally indicate that TL is a promising approach for improving the efficiency of DDMs in building dynamics. However, they are significantly constrained by focusing on individual case studies. Another limitation is that only one specific period of the year was used for fine-tuning. Accordingly, there is no analysis of different fine-tuning periods by season.

As it is essential to demonstrate the technology across various buildings, the authors of \cite{pinto2022sharing} investigated TL on multiple targets. They tested TL for building dynamics on a synthetic building operation dataset, selecting a single source for pretraining. Subsequently, they fine-tuned the model on different targets. In a feature importance analysis, they found that the most important factor for successful TL seems to be the climate.
A limitation of the study is that only one source building was used for pretraining, and only one period of the year (i.e., one season) was considered for fine-tuning. 
Also, the feature importance analysis on building parameter level is limited to only three levels of energy efficiencies.

The authors of \cite{li2024building} investigated different neural net architectures for TL. They found that a CNN-LSTM architecture works slightly better than a plain LSTM.
Furthermore, 3 different source selection methods were investigated. Exhaustive search was found to be best within an archetype of sources having similar building characteristics. Although random selection within the same archetype performed slightly worse, it was preferred due to computational efficiency. As the authors stated, their study is limited to using single sources. Consequently, it remains unclear how random source selection performs in comparison to a multi-source approach. 

In energy demand forecasting, several studies \cite{ribeiro2018transfer, fang2021general, lu2023multi} have proposed multi-source approaches for TL. Their task is different, as only the building load is predicted. 
Studies that use multiple sources for TL are not yet available in the field of building thermal dynamics. Though it was requested by several authors from this community \cite{pinto2022transfer, jiang2019deep, li2024building}. 

Our literature review reveals a research gap in the area of multi-source TL for building thermal dynamics. It also shows that selecting the right source model is still challenging. Further, there is a lack of analysis on how different fine-tuning periods perform by season.
We aim to address this research gap by introducing a general source model that is pretrained on a diverse set of buildings data. Subsequently, the general source model is fine-tuned on a variety of target buildings for each season of the year. Thereby, the step of selecting the right source model is eliminated. This approach will facilitate the widespread use of DDM because only a small amount of data from a target building is required for fine-tuning an existing general source model.\\

The contributions of this paper can be summarized as follows:
\begin{itemize}
    \item We propose the first multi-source approach for TL in building thermal dynamics, presented as a general source model for single-family houses in Central Europe.
    \item The general source model is benchmarked against single-source to single-target TL. We show a smaller mean and variance for the temperature prediction of the fine-tuned model based on the general source model.
    \item We introduce a new fine-tuning evaluation metric to estimate fine-tuning success independently on the target data season. Accordingly, fine-tuning is performed for each season, and the average error is used for evaluation.    
\end{itemize}

The remainder of this paper is organized as follows: in Section \ref{ch:method}, the underlying building simulation is presented. Moreover, the pretraining and fine-tuning of the general source model are explained, followed by the evaluation method. Section \ref{ch:exp} presents the experiments comparing the general source model with the single-source to single-target approach. We discuss our results in Section \ref{ch:disc} and provide a conclusion in Section \ref{ch:conclusion}.

\section{Method}
\label{ch:method}

This section describes the pertaining and fine-tuning of the general source model.
Figure \ref{fig:MethodicalConcept} provides an overview of our approach. First, we generate the time series data with a simulation for the sources and targets as described in Subsection \ref{ch:datagen}. Next, the general source model is pretrained with multiple one-year time series from 450 source buildings (pretraining). Thereafter, the pretrained model is fine-tuned with a small amount of data for individual target buildings (fine-tuning).

\subsection{Building Data}
\label{ch:datagen}

To train and test the general source model, a large amount of time series data from different buildings is required. We produce this data for the sources and the targets through simulations, which run with different building parameters and weather conditions. Simulating the data has the advantage of covering a wide range of buildings with varying insulation values and locations. Additionally, it ensures that building parameters and metadata are explicitly known and can be transparently provided, unlike real-world data, where such information is often unavailable.
For the simulations, we use the Modelica building model from \cite{builDaReview}, which was validated according to the test cases TC900, TC900FF, and TC600FF defined by the standard ANSI/ASHRAE 140-2004 \cite{ASHRAE140_2004}. The model is based on the Modelica Buildings library by \cite{Wetter014buildinglib} and VDI 6007 Part 1 \cite{VDI6007-1}. We customize this model to represent single-family houses in Central Europe. While we demonstrate our method within this specific archetype, the approach can also be applied to other building types. The constraint of the region is based on the climate and the predominant use of heating systems but no cooling systems \cite{schnieders2020design}. TL between different systems is outside the scope of this study since this requires different TL strategies.

\begin{figure}
    \centering
    \includegraphics[width=0.85 \linewidth]{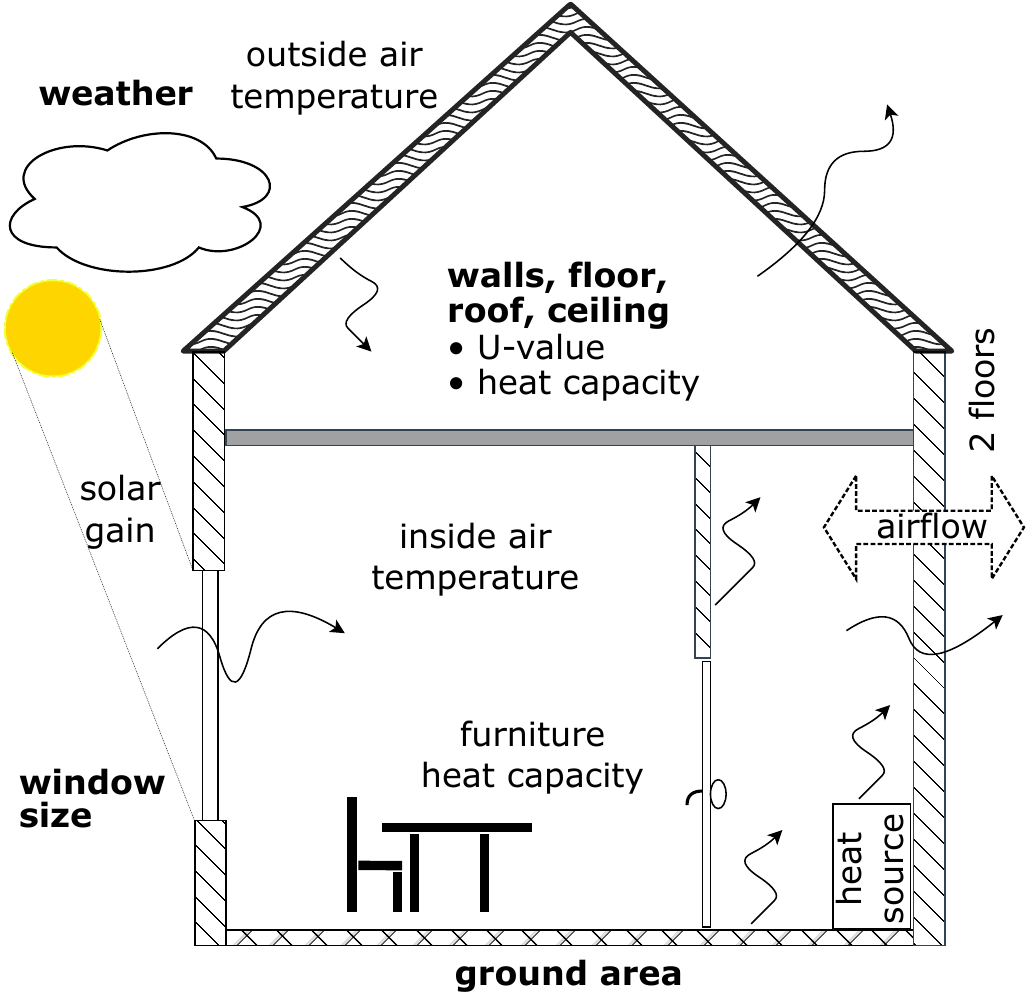}
    \caption{Schematic representation of the Modelica base setup with variable window size, ground area, location, as well as U-value and heat capacity of the envelope}
    \label{fig:building}
\end{figure}

The base setup of the simulation consists of a single-zone building with two floors, a roof, a ceiling, a floor, outside walls, indoor walls, furniture, windows facing each direction, and a heating system (cf. Figure \ref{fig:building}). According to \cite{TASK442013}, this represents a large number of single-family houses. 
The heat source is modeled as ideal, allowing the representation of various heating systems. The heater operates with 50\% radiation and 50\% convection. For each building setup, the simulation calculates the nominal power of the heat source based on transmission and ventilation losses through the building envelope, according to \cite{normHeizlast}. We employ the nominal power to be the maximum power. This ensures that each building is equipped with a properly dimensioned heat source.
The heat source operates with a proportional controller for indoor temperature control. For 30\% of the simulated buildings, a constant setpoint is randomly selected between 20 \textdegree C and 24 \textdegree C. For the remaining 70\%, the simulation randomly selects a constant setpoint for the day in the same range as above, with an additional night setback randomly chosen between 0.5 \textdegree C and 4 \textdegree C. We included scenarios with and without day-night temperature adjustments and selected the setpoint ranges to reflect realistic operating conditions. All ranges were sampled in 0.5 \textdegree C intervals. 

The Modelica model was exported as an FMU and imported in Python using FMPy \cite{FMPy}. The FMU allows for the variation and simulation of buildings with different parameters and weather conditions in Python. 
We vary the base setup in the FMU to create different building simulations. The building parameters selected for variation are those that have the greatest influence on the thermal behavior of the building \cite{Thomas2006Environmental}: the insulation level of the exterior wall ($U$-$value_{wall}$), the area-specific heat capacity of the exterior wall ($c_{wall}$), the size of the building ground area ($A_{ground}$) and the solar gains through windows, i.e., the size of the windows calculated as a ratio to the wall area ($f_{win}$). 
The parameters of the individual building simulations with different values for the source and target buildings can be found in Table \ref{tab:permutations}. The ranges of the parameters were selected according to \cite{tabula}. This way, most of today's single-family houses in Central Europe constructed from 1949 until today can be represented. 
The area-specific heat capacities represent a solid brick, wood stud, and autoclaved aerated concrete wall, as these characterize most of today's buildings with low, medium, and high thermal mass \cite{tabula}. 

We use the weather files from \cite{TMYx} for the building simulation. The choice of locations is based on cities of Central Europe with a cold-temperate climate according to \cite{schnieders2020design}. Here heating is used for the winter and no cooling for the summer. The cities utilized for the source simulations are Belgrade, Prague, Berlin, London, and Zurich.
The targets use weather observed in Munich, Amsterdam, and Bratislava. The cities cover the entire geographical range of the climate zone for both sources and targets.

Each combination of parameters from Table \ref{tab:permutations} is simulated with each weather file for the sources and targets individually. For the sources, this involves 5 U-values, 3 heating capacities, 2 window sizes, 3 ground areas, and 5 locations, resulting in a total of 450 simulations, as indicated in Figure \ref{fig:MethodicalConcept} (S1 - S450). Each time series represents one building with a unique set of parameters and weather. For the targets, we obtain a number of 144 building simulations due to the parameter combinations for the three locations. All simulations run over one year with a 15-minute sampling interval.
We use the generated time series to pretrain and fine-tune the general source model, as explained in the following.

\begin{table}[!b]
    \caption{Parameter values for generating time series}
    \label{tab:permutations}
    \centering
    \begin{tabular}{@{}c|c|c@{}}
        \textbf{Parameter} & \textbf{Sources} & \textbf{Targets} \\ \hline
        $U-value_{\text{wall}}$ [W/(m\textsuperscript{2}K)] & $\{0.1, 0.4, 0.7, 1, 1.3\}$ & $\{0.25, 0.55, 0.85, 1.15\}$ \\ \hline
        $c_{\text{wall}}$ [kJ/(m\textsuperscript{2}K)] & $\{30, 165, 300\}$ & $\{40, 150, 280\}$ \\ \hline
        $f_{\text{win}}$ [-] & $\{0.15, 0.2\}$ & $\{0.16, 0.19\}$ \\ \hline
        $A_{\text{ground}}$ [m\textsuperscript{2}] & $\{60, 90, 120\}$ & $\{70, 100\}$ \\ 
    \end{tabular}
\end{table}

\subsection{Model Pretraining}
\label{ch:train}

The proposed general source model predicts the indoor temperature of the next time steps (output) based on current conditions (weather and temperature) and the control signal.
Therefore the inputs for the general source model consist of the room temperature $T_{in,t}$, the outside temperature $T_{out,t}$, the direct and diffuse solar irradiation $Q_{dir,t}$, and $Q_{dif,t}$, as well as the heat source control signal $u_{in,t}$ (ranging from 0 to 1) of the current time step $t$. A control signal of 1 refers to the maximum power of the heat source (cf. Section \ref{ch:datagen}). We observed no significant differences when using the control signal compared to the heating power as input. However, we found the control signal to be more intuitive since this would be used in reality.
Additional inputs, such as seasonality information (like time of the day), had minimal impact on the prediction performance of the consecutive time steps and were therefore omitted. 

\begin{figure}[h!]
    \centering
    \includegraphics[width=1 \linewidth]{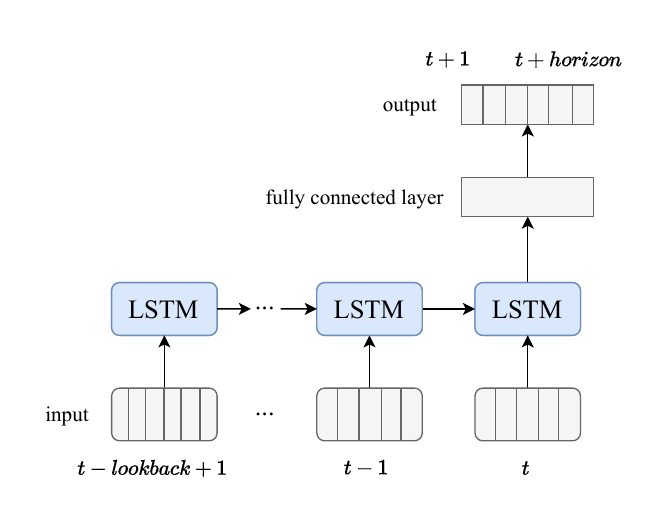}
    \caption{Architecture of the Neural Net}
    \label{fig:architectue}
\end{figure}

An LSTM was chosen for the model due to its ability to capture both long-term (e.g., thermal mass) and short-term (e.g., solar gain) dynamics. An LSTM model has already shown positive results in \cite{pinto2022sharing} for indoor temperature prediction and TL. Our LSTM model includes a fully connected layer to generate an output sequence with the length of the forecast horizon. We mainly use a horizon of 4 hours (16 time steps), which reflects a trade-off between TL literature \cite{chen2020transfer, li2024building, pinto2022sharing}, modeling thermal inertia \cite{thermResponse} and MPC application \cite{Hilliard02072016}. Also, we compare several horizon lengths in Section \ref{ch:exp} since especially the application of MPC requires larger horizons. 
We utilize the sliding window approach, where each subsequent prediction starts from the next time step. The overall architecture is depicted in Figure \ref{fig:architectue}. We used PyTorch for the implementation \cite{paszke2019pytorch}.
Hyperparameter tuning for this architecture was conducted using Bayesian optimization with Optuna \cite{optuna_2019}. In our study, we utilized uniformly distributed hyperparameter values with the results presented in Table \ref{tab:hyperparameter}.

Pretraining of the general source model is based on 450 simulated time series of different source buildings as shown in Figure \ref{fig:MethodicalConcept} (S1 - S450) and Table \ref{tab:permutations} (sources). Employing more buildings would imply sampling the parameters from Table \ref{tab:permutations} more granular as the range is given by \cite{tabula} and incorporating additional locations. This provided only marginal performance improvements while increasing computational demands. Thus, selecting 450 source buildings represented a deliberate trade-off that achieved satisfactory results. 
Before training, the data is pre-processed such that the time series are min-max normalized and subsequently divided into a $70\%$ train, $15\%$ validation (used for hyperparameter tuning), and $15\%$ test set. More specifically, the test and validation set cover September to December, as shown in Figure \ref{fig:MethodicalConcept}. 
The test set only includes winter months and is, therefore, not representative of the entire year. However, the test error during pretraining serves merely as an indication of the prediction quality. Ultimately, the goal is to achieve a low prediction error in the target building after fine-tuning. To evaluate the performance of the general source model, we focus on the test error after fine-tuning.

For pretraining the general source model, the time series of all buildings need to be sliced into training examples. 
We fill each batch with a subset of training examples from a single building, arranged in continuous order (cf. Figure \ref{fig:MethodicalConcept}, top right). Moreover, before each training epoch, we randomly shuffle all batches across all building time series while retaining training examples in continuous order within one batch. This approach leads to slightly better performance in our case than the conventional approach of randomly selecting training examples into batches. 
The reason for this might be the large diversity in the training data of 450 different buildings. Our approach leads to batches that exhibit less variability. Thus, each batch provides a gradient that more steadily points toward the cost minima than a randomly filled batch. The relatively small batch size and random shuffling of batches across epochs possibly result in a regularization effect comparable to the regularizing effect of conventional stochastic gradient descent. However, all experiments were also carried out using the conventional method, which produced similar results.

\begin{table}[!b]
    \centering
    \caption{Hyperparameter selection}
    \begin{tabular}{ c| c | c | c}
         \textbf{Parameter}&  \textbf{Range} & \textbf{Step} & \textbf{Selected}  \\ \hline 
         batch size & $[128-1024]$ & 128 & 256 \\ \hline
         learning rate& $[0.1-1.5]*10^{-3}$ & 0.0001 & 0.0012\\ \hline
         lookback &  $[60-108]$ & $12$ & 96\\ \hline
         LSTM layers& $[1-5]$ &   1 & 3\\ \hline
         neurons per layer& $[75-200]$ &   25 & 125\\ \hline \hline
         optimizer&  &    & Adam \cite{kingma2014adam} \\ \hline
         cost function&  &    & MSE\\
    \end{tabular}
    \label{tab:hyperparameter}
\end{table}

In the next step, the resulting pretrained model can be fine-tuned to a specific target building, as explained in the following subsection.

\subsection{Model Fine-Tuning}
\label{ch:fine-tuning}

Fine-tuning refers to using a pretrained model (source) as a starting point for fitting an ML model to a target.
After the general source model has been pretrained, it can be fine-tuned using data from a target building to capture its dynamics. A major advantage of this approach is data efficiency. Two fine-tuning approaches are particularly popular: weight initialization and layer freezing. With weight initialization, all parameters of the pretrained model are adjusted during fine-tuning. In contrast, layer freezing leaves the parameters of certain "frozen" layers unchanged and only adjusts those of the non-frozen layers. We follow the authors of \cite{pinto2022sharing}, who found that weight initialization performs slightly better for a similar task and architecture. Hence, we use weight initialization for fine-tuning.

To investigate the fine-tuning capability of the general source model, tests are carried out on 144 target time series according to Table \ref{tab:permutations} (targets). The number of targets results from the parameter and weather combinations, as explained in Section \ref{ch:datagen}. Using significantly fewer buildings would have reduced the study's scope, while including more buildings would have increased computational effort. Moreover, the parameters of the target buildings cover a similar range to the parameters of the source buildings. We do this because the general model serves as a pretrained model for this archetype of buildings, i.e., single-family houses in Central Europe.
To avoid having identical buildings for pretraining and fine-tuning, we utilize different parameter values for the source and target buildings. 
Furthermore, we incorporate varying locations for the buildings by using different weather files. This ensures different time series for the sources and targets.
The target buildings represent "in-distribution" fine-tuning data. No "out-of-distribution" buildings are analyzed.

As a first step of fine-tuning, the LSTM parameters of the general source model are used for initialization. Second, the model parameters are fine-tuned, i.e., adjusted with gradient descent on the limited amount of target data. For consistent scaling, the same scaling as in pretraining is applied to the target data. A separate hyperparameter tuning for fine-tuning was not performed. Apart from learning rate and batch size, all hyperparameters depend on the pretrained model.

Different target data lengths are considered for fine-tuning since the length of the fine-tuning data has a major influence on the success of TL \cite{peirelinck2022transfer}. The remaining data of the one-year time series is used as test data. To prevent overfitting on the small fine-tuning set, a best model selection strategy is employed. Best model selection identifies the model with the lowest test error over the training epochs. After training, we use the best model for evaluation.

The code for pretraining and fine-tuning is available on GitHub \cite{GenTLGit}.
For training, we used an AIME T600 workstation equipped with an NVIDIA RTX A6000 GPU (48 GB), a Threadripper Pro 5995WX CPU (64 cores, 2.7 / 4.5 GHz), and 512 GB of RAM. Pretraining took 3 hours and 43 minutes,  4.28 GB of GPU memory, and 46,5 GB of system memory. Fine-tuning for one target required at most 1.2 GB of GPU memory, 1.3 GB of system memory, and 56 seconds of training time. In Section \ref{ch:exp}, we will execute experiments that compare the general pretrained model with single-source models during fine-tuning. Depending on the hyperparameter configuration, the single-source models required between 21 seconds, and 2 minutes and 36 seconds for fine-tuning. 97\% of the fine-tuned single-source models that arose from the experiments needed less than one minute. The resources used for single-source fine-tuning are the same as those used for fine-tuning the general source model.

\subsection{Evaluation}
\label{ch:eval}

For evaluation, the general source model is fine-tuned on various targets. For this purpose, the fine-tuned prediction model is compared to the ground truth of the target. We use the mean absolute scaled error (MASE) as error metric. The MASE compares the absolute forecast error (numerator) with the absolute error of a naive predictor (denominator). We use the value of the output variable in the last time step as naive predictor (persistence-based forecast). Thus, the MASE reflects the general difficulty of a prediction task. This is especially useful in slow-changing building dynamics, where short-term forecasting is rather simple. 
In addition, the root mean square error (RMSE) is used, as it represents the error in degrees Celsius, whereas the MASE provides a ratio. Both formulas are shown below for a horizon of h time steps:
\begin{equation}
\begin{split}
    \text{MASE} &= \frac{\frac{1}{n} \sum_{i=1}^{n} \frac{1}{h} \sum_{j=1}^{h} |T_{in,i+j} - \hat{T}_{in,i+j}|}{\frac{1}{n} \sum_{i=1}^{n} \frac{1}{h} \sum_{j=1}^{h} |T_{in,i+j} - T_{in,i}|} \\
    \text{RMSE} &= \sqrt{\frac{1}{n} \sum_{i=1}^{n} \frac{1}{h} \sum_{j=1}^{h} (T_{in,i+j} - \hat{T}_{in,i+j})^2}
\end{split}
\end{equation}

TL in the buildings domain typically assumes that only a few days of target data are available. The effectiveness of TL is therefore influenced by the season from which this data originates (e.g., summer or winter) \cite{pinto2022transfer}. In the following, we refer to the seasonal origin of the target data as the collection period. As discussed in Section \ref{ch:review}, existing studies have only evaluated their TL models using one collection period in the target. However, a pretrained model should be evaluated several times for fine-tuning with different collection periods. The aim is to evaluate scenarios where fine-tuning data is collected during certain seasons and not to create season-specific models. 

We introduce a new approach to assess the success of fine-tuning and to avoid any influence from the collection period (see Algorithm \ref{alg:fine-tune}). 
According to the seasons in Central Europe, we consider four different collection periods ($seasons$) in the year for fine-tuning: winter, spring, summer, and fall. The pretrained model ($pretrained\_model$) is fine-tuned separately for each season with the chosen data length ($data\_length$), resulting in four different models for each target. Thus, each fine-tuned model is representative of a specific collection period in the target building. We use the remaining data of the year ($test\_data$) for testing. The prediction error ($error\_ft\_Mi$) is calculated for each fine-tuned model ($ft\_Mi$). As each collection period leads to a different fine-tuned model, the prediction performance varies on the test data. These errors are then averaged to determine the overall error ($mean\_error\_ft\_M$). Compared to the evaluation of a fine-tuned model on a single period, the overall error ($mean\_error\_ft\_M$) provides a more robust estimation as it assesses the performance across multiple seasons.

\begin{algorithm}
    $seasons \gets 4$ \;
    $days \gets data\_length$ \;
    $M \gets pretrained\_model$ \;
    $offset \gets 365/seasons $ \;
    \For{i in range (0, seasons-1)}{
        $ft\_data \gets target\_data[i*offset: i*offset+days]$ \;
        $test\_data = target\_data-ft\_data$ \;
        $ft\_Mi \gets finetune \: M \: on \: ft\_data$ \;
        $error\_ft\_Mi \gets test \: ft\_Mi \: on \: test\_data$ \;
        $results.add(error\_ft\_Mi)$ \;
    }
    $mean\_error\_ft\_M \gets results / seasons$ \; \
    \caption{Fine-tuning evaluation metric}
    \label{alg:fine-tune}
\end{algorithm}

\section{Experiments}
\label{ch:exp}

This section illustrates the experiments conducted to evaluate the general source model described in Section \ref{ch:method}. Therefore we first present a plot over time to give an intuition of the prediction result. The remaining experiments serve to compare the fine-tuned general source model with the conventional approach of single-source to single-target TL. Therefore we first display the experimental setup in Section \ref{ch:ex_setup} and then the results in Section \ref{ch:results}. 
\\ \\
\begin{figure}
    \centering
    \includegraphics[width=1 \linewidth]{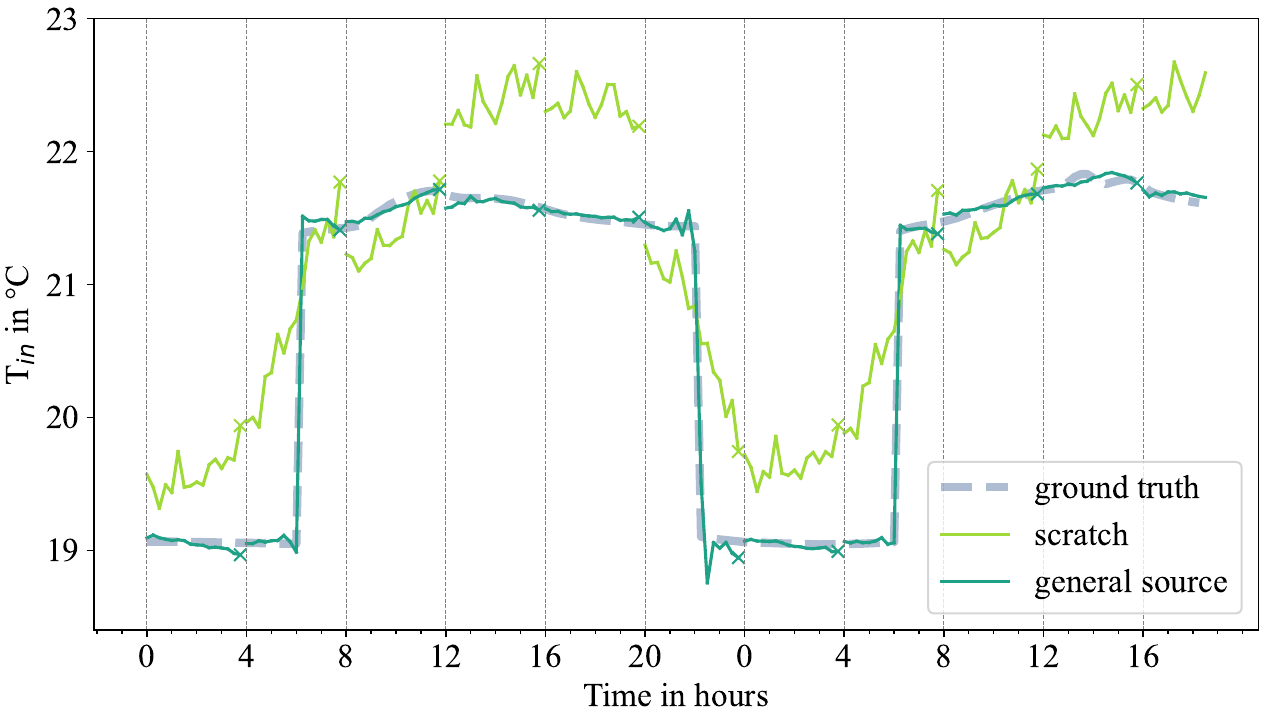}
    \caption{Indoor temperature prediction of a model trained from scratch and a model fine-tuned on the general source model}
    \label{fig:final_plot_over_time}
\end{figure}
In the plot over time, we compare the TL result with a model trained from scratch and the ground truth. We utilize target building T6 from Table \ref{tab:ss2st} and 30 days of available data as an example. This data is used to fine-tune the general source model as described in Section \ref{ch:fine-tuning}. Additionally, the data is used to train an ML model from scratch, which means the model is initialized without pretrained parameters. Training an ML model from scratch is the conventional approach when no TL is applied. 
We use the same architecture as in Figure \ref{fig:architectue} with separate hyperparameter tuning for the model from scratch. As shown in other studies, LSTMs are a suitable choice for this task \cite{Elmaz2021CNNLSTM, Martinez2020LSTMMLP, Mtibaa2020LSTM}.

Figure \ref{fig:final_plot_over_time} presents the results. Only forecast horizons shifted by 16 time steps are plotted instead of all sliding windows. The end marker ("x") and a dashed vertical line indicate the horizon. 
Although the plot delivers only anecdotal evidence for a particular target building, it provides some intuition about the effectiveness of TL. The model trained from scratch follows the general trend but exhibits some instabilities and larger deviations. This is because the ML model had too little data available for training. In contrast, the fine-tuned model predicts the temperature very well. However, slight deviations from the ground truth are visible.

\subsection{Experimental Setup}
\label{ch:ex_setup}

This section outlines the design of the single-source to single-target approach and the experimental setup for comparison with the general source model.

For the single-source approach, we follow \cite{pinto2022sharing}. We use two years of data from one source building to pretrain the architecture shown in Figure \ref{fig:architectue}. Each single-source model undergoes its own hyperparameter tuning. This ensures the optimal LSTM configuration for each source building. The train, validation, and test split is 70\%, 15\%, and 15\%, respectively.
For fine-tuning the single-source models, the same procedure and evaluation as for the general source model are applied (cf. Section \ref{ch:fine-tuning} and \ref{ch:eval}). 
\\ \\
In the following, we outline the experiments and their setup. In each experiment, we analyze the prediction error ($mean\_error\_ft\_M$) as presented in \ref{ch:eval}. The results will be shown in Section \ref{ch:results}. 

We begin with a \textit{small-scale} analysis to assess the performance of several single-source models and the general source model for fine-tuning individual target buildings. This experiment allows a detailed comparison of the fine-tuning results for specific targets. A subset of 10 sources and 8 targets was selected from Table \ref{tab:permutations}, with the corresponding results provided in Table \ref{tab:ss2st}. The sources and targets were selected so that a representative sample from the overall distribution was ensured. This selection includes both highly similar and highly different source-to-target pairs regarding building characteristics and geographical location. 30 days of target data were utilized for fine-tuning the single sources and the general source model. Thus, 11 pretrained models were fine-tuned per target, resulting in 88 models. Subsequently, all fine-tuned models are examined for their prediction accuracy and compared with each other.

The second experiment serves to investigate different forecasting horizons. For real-world deployment, different horizons might be necessary. A small horizon was used in related TL Literature \cite{chen2020transfer, li2024building, pinto2022sharing} whereas larger ones are more applicable to MPC deployment \cite{Hilliard02072016}.
Therefore, we compare the fine-tuned general source model to fine-tuned single-source models for a 1, 4, 12, and 24 hour forecast horizon. We use the same sources and targets as in the previous experiment. Pertaining has to be performed for the general source and the single sources for each horizon.

In the third experiment, we analyze the effect of different target data lengths for fine-tuning. Similar to \cite{pinto2022sharing, chen2020transfer}, we consider 10, 30, and 60 days of target data lengths. This analysis evaluates how the amount of fine-tuning data influences the success of TL for single-source models compared to the general source model. We also assess the impact of data length for training a model from scratch, allowing for comparison with TL models. The model trained from scratch follows a similar design to the one in the plot over time. We use the fine-tuning data from the target for training. The remaining data of the year serves as test data, as in Section \ref{ch:eval}. We apply separate hyperparameter tuning for each model trained from scratch. 

Finally, we perform a \textit{large-scale} analysis to provide broader insights into the applicability of the method. In this analysis, we compared the distributions of prediction errors for multiple targets fine-tuned on single-source models with those fine-tuned on the general source model. The experiment also examines the performance of the general source model compared to random source selection. We sample 500 randomly selected single-source to single-target pairs for this analysis. The sources are sampled from Table \ref{tab:permutations} (sources), and models are pretrained as described in \ref{ch:ex_setup}. The targets are sampled from Table \ref{tab:permutations} (targets) and used for fine-tuning. The general source model is fine-tuned on all of the 144 targets. For fine-tuning 30 days of target data are used. 

The subsequent section provides the results of the experiments.

\begin{table}[!b]
    \centering
    \caption{List of single sources (S1-S10) and single targets (T1-T8)}
    \begin{tabular}{c|c|c|c|c|c|c|c}
        & \rotatebox{90}{$U$-$value_{wall}$ $[W/(m^2K)]$} & \rotatebox{90}{\textbf{$c_{wall}$ $[J/(m^2K)]$}} & \rotatebox{90}{\textbf{$f_{win}$}}  & \rotatebox{90}{\textbf{$A_{ground}$ $[m^2]$}} & \rotatebox{90}{$T_{sp,day}$ [\textdegree C] } & \rotatebox{90}{$\Delta T_{night}$ [\textdegree C]} & \rotatebox{90}{Weather} \\  \hline
        S1 & 0.10 & 165,000 & 0.15 & 120 & 22.5 & 0.0 & Zurich \\ \hline
        S2 & 0.10 & 300,000 & 0.20 & 60 & 24.0 & 2.5 & Belgrade \\ \hline
        S3 & 0.40 & 165,000 & 0.20 & 120 & 21.0 & 3.5 & Prague \\ \hline
        S4 & 0.40 & 30,000 & 0.15 & 90 & 24.0 & 3.0 & Berlin \\ \hline
        S5 & 0.70 & 30,000 & 0.15 & 90 & 22.5 & 0.5 & London \\ \hline
        S6 & 0.70 & 300,000 & 0.20 & 60 & 20.5 & 0.0 & Zurich \\ \hline
        S7 & 1.00 & 165,000 & 0.20 & 120 & 22.0 & 3.5 & Berlin \\ \hline
        S8 & 1.00 & 30,000 & 0.15 & 60 & 21.0 & 3.5 & Prague \\ \hline
        S9 & 1.30 & 30,000 & 0.15 & 90 & 21.0 & 2.0 & Belgrade \\ \hline
        S10 & 1.30 & 300,000 & 0.20 & 120 & 23.5 & 1.5 & London \\ \hline \hline
        T1 & 0.25 & 280,000 & 0.19 & 100 & 21.0 & 0.0 & Amsterdam \\ \hline
        T2 & 0.25 & 40,000  & 0.16 & 70 & 22.0 & 1.0 & Bratislava \\ \hline
        T3 & 0.55 & 150,000 & 0.16 & 70& 23.0 & 3.0 & Amsterdam \\ \hline
        T4 & 0.55 & 280,000 & 0.19 & 100 & 20.5 & 1.5 & Munich \\ \hline
        T5 & 0.85 & 150,000 & 0.19 & 100 & 22.5 & 0.5 & Bratislava \\ \hline
        T6 & 0.85 & 40,000 & 0.16 & 70 & 22.0 & 2.5 & Munich \\ \hline
        T7 & 1.15 & 280,000 & 0.16 & 70 & 23.0 & 0.0  & Bratislava \\ \hline
        T8 & 1.15 & 40,000 & 0.19 & 100 & 23.0 & 1.5 & Amsterdam \\
     \end{tabular}
     \label{tab:ss2st}
\end{table}

\subsection{Results}
\label{ch:results}

\begin{figure*}
    \centering
    \includegraphics[width=0.75 \linewidth]{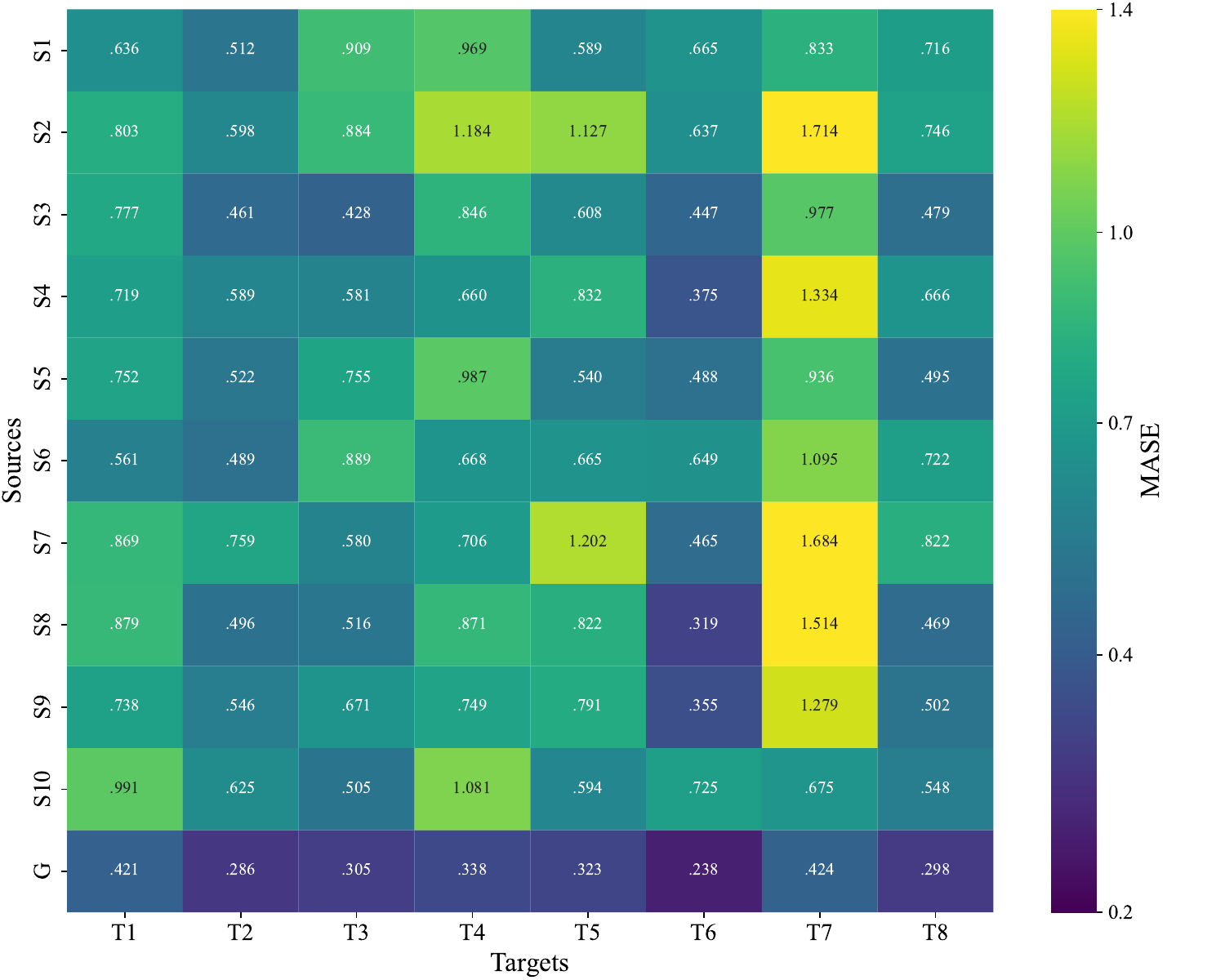}
    \caption{Heat map of the prediction performance expressed as the mean absolute scaled error (MASE) of target buildings fine-tuned on different single sources (S1 to S10) and the general source model G, respectively}
    \label{fig:heat_map}
\end{figure*}

The first experiment demonstrates the \textit{small-scale} analysis. The results of this experiment are shown as a heat map in Figure \ref{fig:heat_map} (RMSE values are detailed in the appendix Figure \ref{fig:heat_map2}). The plot shows the targets 1 to 8 (horizontal axis) fine-tuned on the source models 1 to 10 as well as on the general source model (vertical axis). The value and the corresponding color represent the MASE. The general source model has the lowest MASE across all targets shown. The MASE for the general source model ranges between a minimum value of 0.238 and a maximum of 0.424. In comparison, the MASE of the single sources is between 0.319 and 1.714, indicating overall higher errors and variance of MASE values. 
Moreover, a different single source performs best for each individual target. For example, S8 works best for the target building T6 but performs poorly on T1. S6, on the other side, works best among the single sources for T1 but leads to an error on T6 that is twice as high as its best. In some cases, random source selection could lead to negative transfer, as the MASE exceeds 1. This indicates that the TL model performs worse than the naive predictor. In the heat map, this is the case for all single sources for at least one target, except for S1, S3, and S5. In contrast, the fine-tuned general source models always exceed the naive predictor.
\\ \\
\begin{figure*}
    \centering
    \includegraphics[width=1 \linewidth]{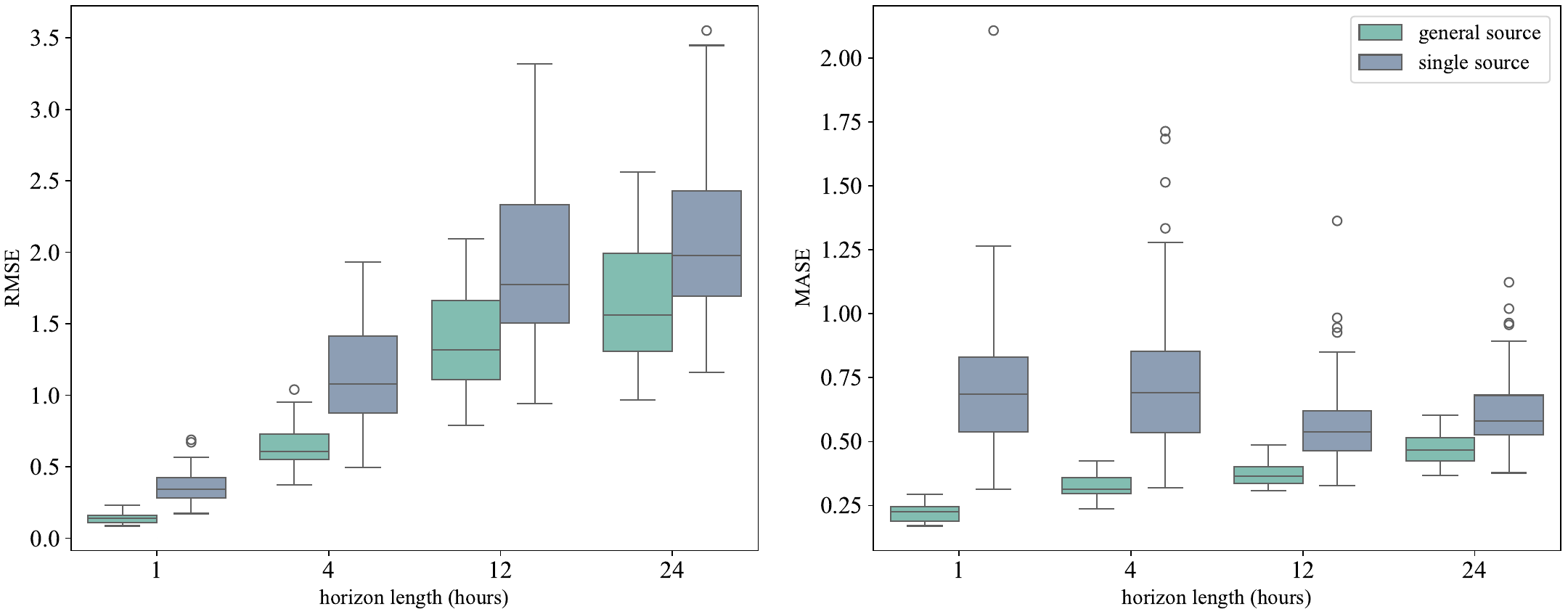}
    \caption{Box plots of fine-tuned general source and single sources for different forecast horizons shown for root mean squared error (RMSE) and Mean absolute scaled error (MASE) values}
    \label{fig:horizons}
\end{figure*}
The second experiment compares different forecast horizons. We repeat the previous experiment for each horizon and present the distributions of the fine-tuned single sources and the fine-tuned general source as a box plot. Figure \ref{fig:horizons} depicts the RMSE and MASE values. The box plots display the 25th percentile (first quartile), 50th percentile (median), and 75th percentile (third quartile). Whiskers extend to 1.5 times the interquartile range from the first and third quartiles, with outliers indicated as values beyond this range. The RMSE increases for all models with an increase in the forecasting horizon since the prediction task gets more challenging. Additionally, the spread of the whiskers and quartiles widens, indicating increased variance.
The MASE values evolve differently from the RMSE values as the horizon length increases because they compare the LSTM's performance to the naive predictor. The MASE of the fine-tuned general source models trends slightly upward with an increase in horizon length. The fine-tuned single sources, on the other hand, trend downwards with median and variance. This might be due to increasing prediction errors of the naive predictor. The MASE of the fine-tuned single sources has the lowest median for 12 hours. 
A reason for this could be that after 12 hours, depending on the starting point and the day-night setback, some buildings have different temperature setpoints while others don't (cf. Section \ref{ch:datagen}). However, after 24 hours, all buildings have again the same setpoint.

There seems to be some indication of convergence of the RMSE and MASE distributions of the fine-tuned general source model and the fine-tuned single-source models for larger forecast horizons. This might be due to the stochastic nature of weather (solar gains, outside temperature) which becomes the dominant cause of the variance in forecast errors as the forecast horizon increases. However, for all forecast horizons, the fine-tuned general source model has lower medians, interquartile ranges, and spreads of whiskers for the RMSE and MASE values compared to the fine-tuned single-source models.
\\ \\
\begin{figure*}
    \centering
    \includegraphics[width=1 \linewidth]{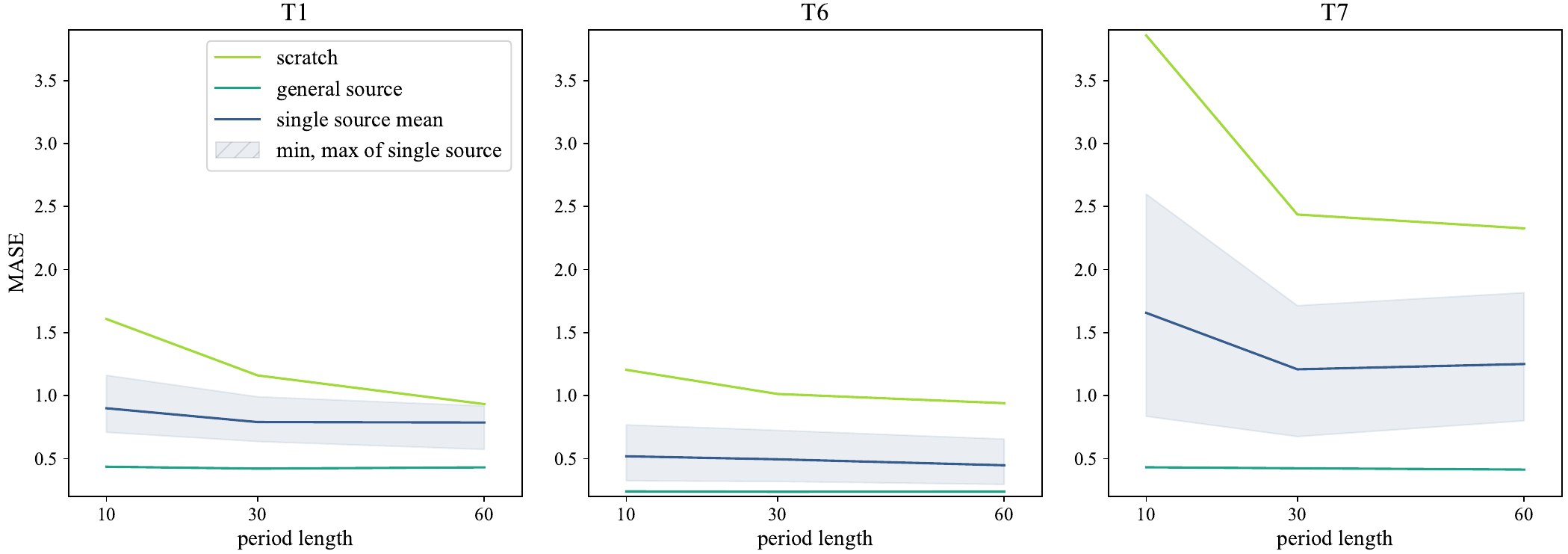}
    \caption{Mean absolute scaled error (MASE) for different training approaches and fine-tuning periods (10, 30, and 60 days)}
    \label{fig:final_period_length_mase}
\end{figure*}
The third experiment evaluates different target data lengths. We use 3 different targets for showcasing. Targets T1, T6, and T7 were selected for their distinct patterns in the heat map from Figure \ref{fig:heat_map}. T7 showed poor results for most sources, while T6 performed best and T1 performed medium. 
The results of the experiment are shown in Figure \ref{fig:final_period_length_mase}. For each target, the x-axis represents the period length. The y-axis shows the MASE for both the fine-tuned models and the models trained from scratch, corresponding to their respective amounts of target data. The gray-blue area and the blue line indicate the spread of the 10 fine-tuned single sources (min, max) and the mean, respectively.
In general, all curves exhibit a downward slope as the training data length increases. This is expected as more training data tends to lead to lower prediction errors. 
For the TL models, the MASE and the variance typically decrease as more data is used for fine-tuning.
However, random shuffling after each epoch can lead to different parameter updates, causing variations in the prediction performance of the models. 

Models trained from scratch struggle to surpass the naive predictor, only doing so with 60 days of data for targets T1 and T6. Target T1 and T6 appear easier to predict in contrast to target T7, as indicated by lower errors across all models. The fine-tuned single-source models always exceed the model from scratch. They also struggle to outperform the negative predictor, especially for T7, and when fine-tuning with small amounts of data for target T1. This is not observed for the fine-tuned general source model, as it consistently outperforms the naive predictor and the other models across all cases.
\\ \\
\begin{figure}
    \centering
    \includegraphics[width=1 \linewidth]{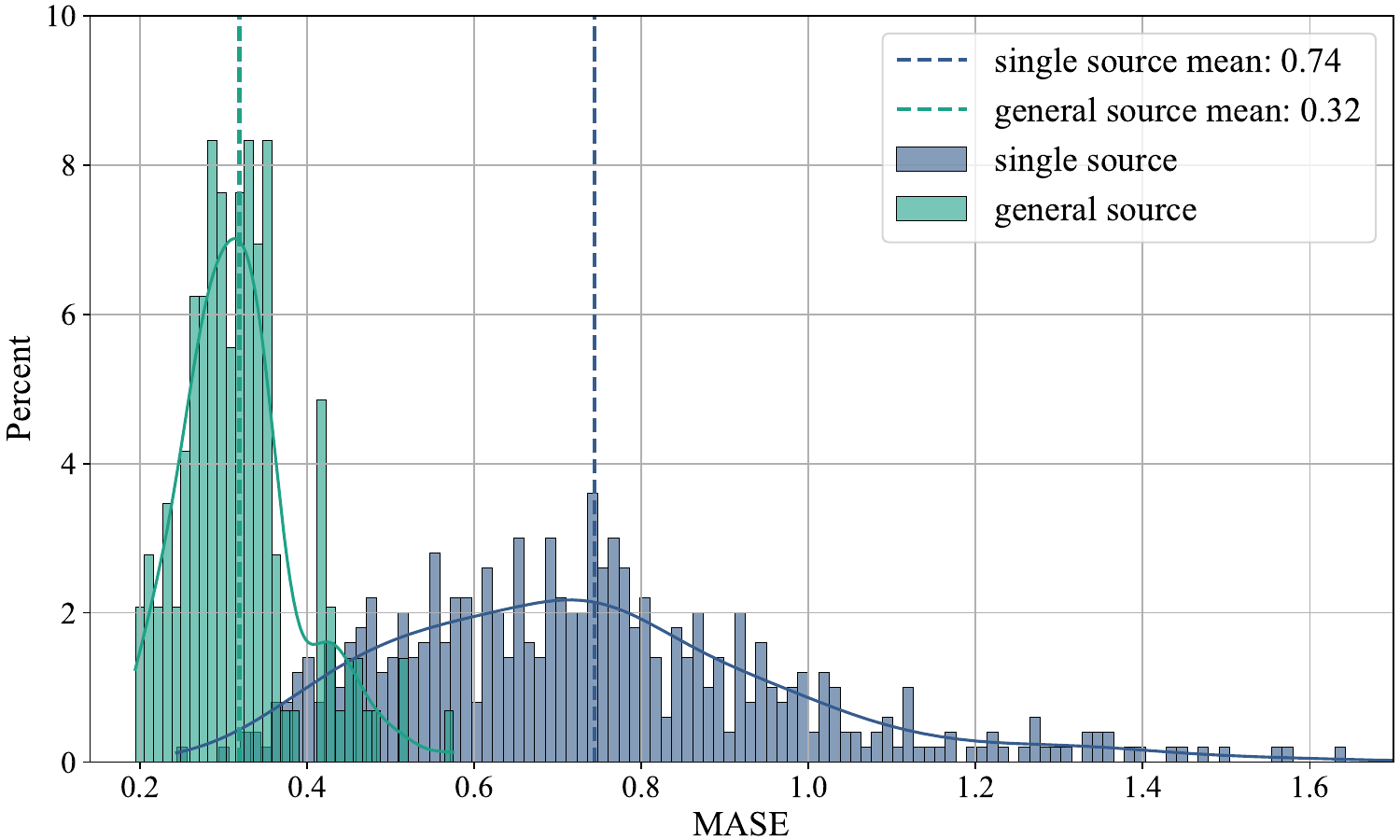}
    \caption{Histograms of MASE distributions for fine-tuned models using single sources (blue) and the general source model (green)}
    \label{fig:histogram}
\end{figure}
The final experiment serves as a \textit{large-scale} comparison for the single sources and the general source model. 
Figure \ref{fig:histogram} shows a histogram of the experiment (RMSE values are detailed in the appendix Figure \ref{fig:histogram2}). The distribution of the MASE values across the single sources (blue) and the general source model (green) after fine-tuning are shown. The general source model exhibits a lower mean and variance compared to the MASE distribution of fine-tuned single-source models. Hence, the performance, as well as the reliability of the general source model approach, seems superior to the single-source approach. For 64 single-source to single-target pairs, the fine-tuning error exceeds 1, indicating negative transfer. The general source model achieves a maximum MASE of 0.58, consistently outperforming a naive predictor. These findings demonstrate the issues of random source selection for single-source to single-target TL compared to the general source model approach. 

\section{Discussion}
\label{ch:disc}

This paper presents a new TL approach for building thermal dynamics based on a general source model. We trained a general source model for the archetype of single-family houses in Central Europe. 450 different time series representing buildings with various construction materials, sizes, and weather conditions are employed. A modifiable FMU-based simulation was utilized for this purpose. The resulting general source model served as the foundation for TL. A small amount of data (10, 30, and 60 days) from different target buildings was used to fine-tune the general source model. The results were compared with the traditional single-source to single-target approach. For fine-tuning, a new evaluation metric was proposed that evaluates the prediction error after fine-tuning for different periods of the year.

Among the studies performed, the general source model appears to be a more reasonable choice for TL compared to single-source TL. In the \textit{small-scale} analysis, no single-source model outperformed the general source model. The \textit{large-scale} analysis revealed that the mean MASE of the fine-tuned general source models was lower by 56.7\% compared to the fine-tuned single sources. Furthermore, none of the experiments demonstrated negative transfer for the general source model.
This enables us to provide clear answers to the earlier stated research questions:

Firstly, we wanted to explore how knowledge from multiple sources can be leveraged for successful TL. It has been shown that the approach of pretraining a general source model is an effective solution. %
Slicing and shuffling the time series from multiple sources while keeping the training examples in continuous order within a batch provides good generalization during pretraining.

Secondly, the study investigates the prediction error of a multi-source pretraining approach compared to single-source methods. Our results demonstrate that the multi-source approach not only significantly reduces prediction errors (MASE and RMSE) on average but also exhibits lower variance across different targets. 

Lastly, this study compares the efficacy of the general source model versus random source selection on an archetype of buildings, i.e., single-family houses in Central Europe. Our findings indicate that random single-source selection can lead to inconsistent performance across various targets, sometimes resulting in negative transfer. In contrast, the general source model performed significantly better on all targets. 
This shows that a general pretrained model as a source should rather be used for fine-tuning than a random single source.

As this study is the first to explore the multi-source approach for building thermal dynamics, there are inherent limitations and open questions.
The study's limitations include the use of synthetic data. Real data from diverse single-family houses in Central Europe with varying insulation properties, sizes, locations, and temperature reference values but identical heating setups, to the best of our knowledge, is currently unavailable. Therefore, elaborating a suitable real-world data set and applying it to the proposed method would be valuable for future research. Addressing this issue would pave the way for the practical implementation of the proposed method. 
Another limitation of this work is the focus on single-family houses from Central Europe. If the data is available, this methodology should be transferable to other building types. Further research is needed to determine whether to develop a foundation model encompassing all building types or to categorize buildings into specific subtypes for developing general models, like in this paper. 
For the general source model, we employed a 3-layer LSTM. Despite its relatively simple architecture, this approach yielded promising results. However, future plans include exploring other architectures, such as transformers, to potentially enhance performance.

\section{Conclusion}
\label{ch:conclusion}

In this paper we present GenTL, a general transfer learning (TL) model for building thermal dynamics.
GenTL was trained on 450 simulated time series representing different single-family houses in Central Europe. In this way, we realize a general source model that enables data-efficient fine-tuning on a wide range of target buildings. The proposed method has been compared to single-source to single-target TL. We found that the general source model results in significantly lower prediction errors on average and smaller variance after fine-tuning.
Through the developed methodology, we hope to advance research in the field of data-driven modeling for energy-efficient control and fault detection \& diagnosis.

\begin{acks}
We thank Markus Wirnsberger, Ferdinand Sigg, and Dominik Aimer for their support in the building simulation. Additionally, we are grateful to Felix Koch and David Broos for their assistance with the project programming.
\end{acks}
\balance
\bibliographystyle{ACM-Reference-Format}
\bibliography{sample-base}
\appendix

\section{Appendix}

\begin{figure*}
    \centering
    \includegraphics[width=0.7 \linewidth]{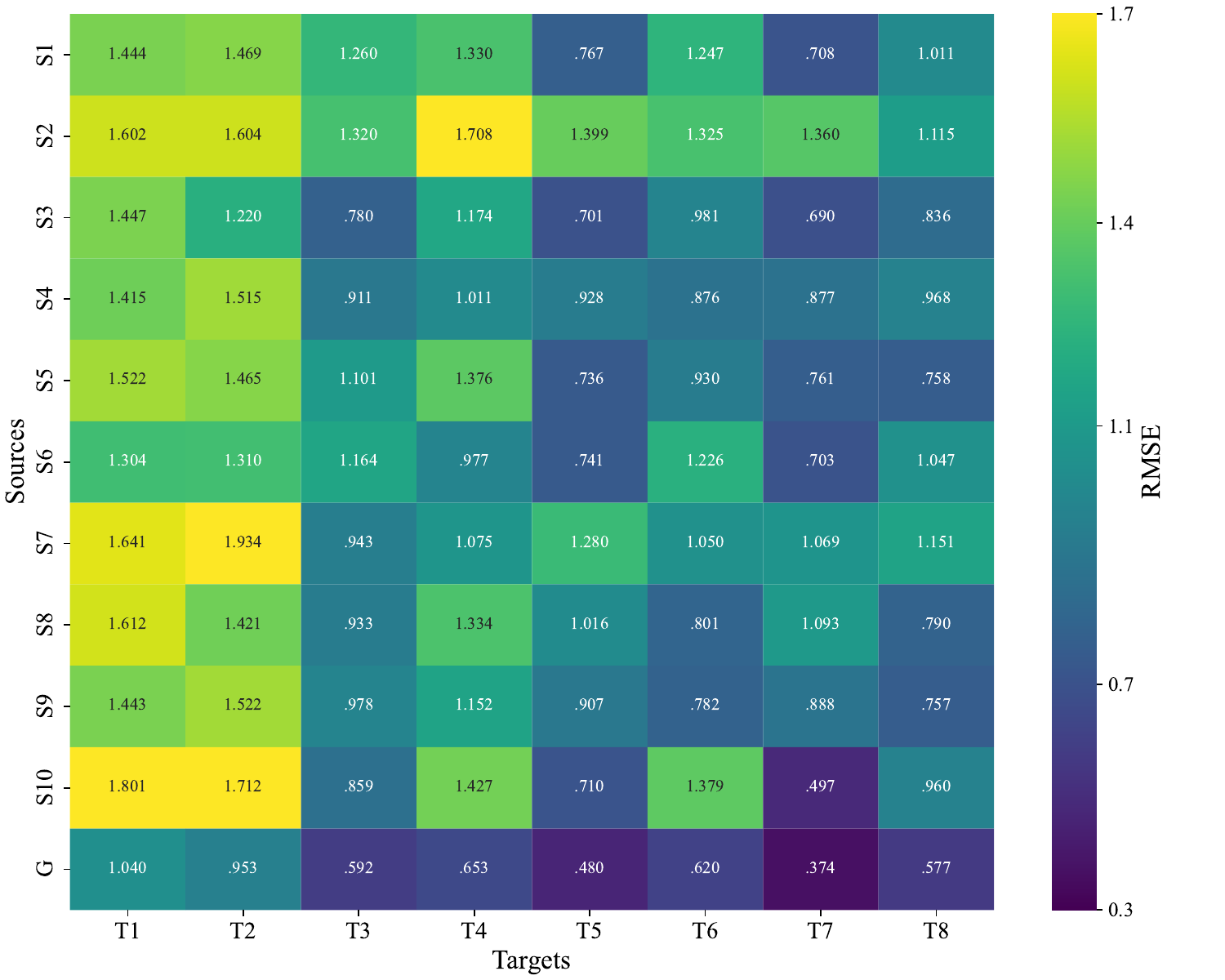}
    \caption{Heat map of the prediction performance expressed as the root mean square error (RMSE)  of target buildings fine-tuned on different single sources (S1 to S10) and the general source model G, respectively}
    \label{fig:heat_map2}
\end{figure*}

\begin{figure*}
    \centering
    \includegraphics[width=0.65 \linewidth]{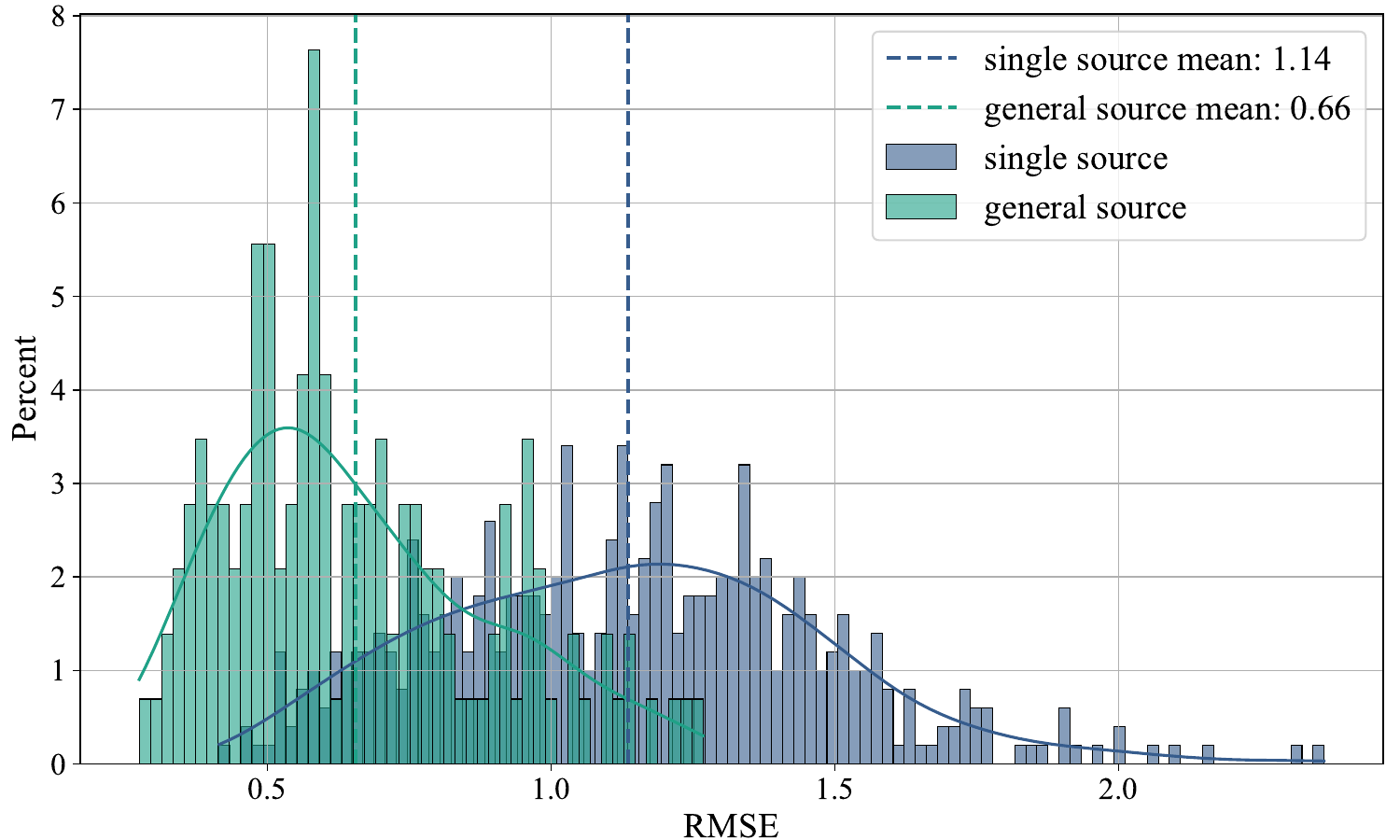}
    \caption{Histograms of RMSE distributions for fine-tuned models using single sources (blue) and the general source model (green)}
    \label{fig:histogram2}
\end{figure*}
\end{document}